\documentclass[12pt]{article}
\usepackage[english]{babel}

\begin{document}
\title{Inverse $\beta^+$ - decay of proton in the presence of  strong magnetic field}

\author{I. Mamsourov $^1$  , H. Goudarzi$^{1,2}$\footnote{E-mail: Goudarzia@phys.msu.ru} \\ {\footnotesize \textit{1. Department of Theoretical Physics, Moscow State University, Moscow 119899, Russia}}\\ \textit{{\footnotesize 2. Department of Physics, Urmia University, Urmia, Iran}}}

\date{}
\maketitle



\begin{abstract}
In this work the energy spectrum and solution of Dirac's equation for the charged and neutral fermions taking into account interaction of the anomalous magnetic moments of particles with the external field are obtained. The total probability of the inverse beta- decay of proton in the presence of strong uniform magnetic field taking into account the anomalous magnetic moments of nucleons is found.
\\
\\
\end{abstract}

\begin{center}
\title\textbf{1. Introduction}
\maketitle
\end{center}

In the present work, inverse $\beta^+$  decay of proton in the presence of the strong uniform magnetic field is studied: ${p^+\rightarrow{n+{e^+} +\nu}}$ , here  $p^+$ , $n$ , $e^+$ and $\nu$ are used for proton, neutron , positron and  neutrino respectively. This process becomes energetically permitted only upon consideration of interaction of anomalous magnetic moments(AMM) of nucleons with the external magnetic field. There is special interest in this process for astrophysics, since the conditions suitable for its flow occur in the neutron stars, on surface of which the magnetic field can reach $10^{15}$ G.

Direct $\beta^- $ - neutron decay in the strong magnetic field without taking into account AMM fermions was examined earlier (for example, see [1]). The so-called  URCA - processes
 
${n\rightarrow{p^+  +{e^ -} + \tilde{\nu}}}$ , ${p^+ + e^- \rightarrow{n+ \nu}}$ , \\ where $e^-$ - electron, $\tilde{\nu}$ - antineutrino, as they assume (for example, see [2]), chemical equilibrium in the degenerate ideal gas of nucleons and electrons, by which, in the first approximation, they simulate substance in the central region of neutron star, is supported.

 In [3,4]  the influence of the ultra frozen magnetic field in  the neutron star to the conditions of chemical equilibrium and the equation of state of the degenerate gas of nucleons and electrons taking into account the contribution of interaction AMM of nucleons with the external magnetic field was studied.
 
At first we will obtain the solutions of Dirac's equation taking into account interaction AMM of fermions with the external magnetic field, which are  the eigenfunctions of Dirac's Hamiltonian with AMM of particles and operator of the polarization,  $\hat\mu_3$. It should be noted that these problems partially were considered in the works [5,6].

  We select system of the units, where \ $\hbar=c=1$.\\
  
\begin{center}
\title\textbf{2. Bispinors and energy spectrum of fermions taking into account AMM}
\maketitle
\end{center}

Dirac's equation for the fermion with AMM in the presence of stationary and uniform magnetic field takes the form:
\begin{equation}
i\partial_t\Psi=\hat{H}_D\Psi.
\end{equation}
where : $$\hat{H}_D=(\vec{\alpha}\cdot\vec{P})+m\rho_3+ \textit{M} \rho_3(\vec{\Sigma} \cdot \vec{H})\;.$$

\begin{equation}
\vec{\alpha}=\rho_1\vec{\Sigma}\; , \  \   \rho_1=\left(\begin{array}{cc}
0 & I \\
I & 0
\end{array}\right) \;, \ \ 
\rho_3=\left(\begin{array}{cc}
I & 0 \\
0 & -I
\end{array}\right)\; , \ \ 
\vec{\Sigma}=\left(\begin{array}{cc}
\vec{\sigma} & 0 \\
0 & \vec{\sigma}
\end{array}\right)\;.
\end{equation} \\
where $\vec{\alpha}$ and $\rho_{1,3}$ - are the Dirac's matrices  and $\sigma_{1,2,3}$ - respectively are Pauli's matrices, and \\$\vec{P}=-i\vec{\nabla}-e\vec{A}$ - is the generalized momentum of fermion, $\vec{A}$ - is the vector potential of electromagnetic field, $m$ - is mass and $e$ is charge, $\textit{M}$ - is the anomalous magnetic moment of fermion and $\psi(\vec r, t)$ - is four-component spinor.

Let us consider the motion of the charged or neutral fermion with AMM in the stationary and uniform magnetic field, described by vector potential of $\vec{A}$ in the following calibration :
\begin{equation}
A_x=A_z=0  , \ \ \ \ \  A_y=Hx .
\end{equation} 

Bispinor $\Psi(\vec{r},t)$ appears in the form:
$$\Psi(\vec{r},t)=e^{-i\varepsilon t}\psi(\vec{r})\;.$$ 
Hence for $\Psi(\vec{r})$  we obtain the following system of equations :
$$(\varepsilon -(m + \textit{M} H))\psi_1-(P_1-iP_2)\psi_4-P_3\psi_3=0$$
$$(\varepsilon -(m - \textit{M} H))\psi_2-(P_1+iP_2)\psi_3+P_3\psi_4=0$$
\begin{equation} 
(\varepsilon +(m + \textit{M} H))\psi_3-(P_1-iP_2)\psi_2-P_3\psi_1=0
\end{equation}
$$(\varepsilon +(m -\textit{M} H))\psi_4-(P_1+iP_2)\psi_1+P_3\psi_2=0.$$ 

let us examin at first the case of charged  fermion(proton). The wave function  $\psi{(\vec{r})}$ we will find  in the following form:
$$\psi{(\vec{r})}=\frac{1}{L}\exp{(ip_2 y+ip_3 z)}f(x)\; .$$
\begin{equation}
f(x)=\left(\begin{array}{c}
C_1\; u_{n-1}(\eta) \\
iC_2 \;u_{n}(\eta) \\
C_3\; u_{n-1}(\eta) \\
iC_4 \;u_{n}(\eta) 
\end{array}\right)\ ; \ \ \  \eta=\sqrt{2\gamma}\; x+\frac{p_2}{\sqrt{2\gamma}}\  , \ \ \  \gamma=\frac{eH}{2}\;.
\end{equation} \\
Here $u_n(\eta)$  - denotes Hermite's function.

Substituting equ.(5) in equ.(4), we will obtain the system of linear equations for the coefficients  $C_i$:
$$(\varepsilon \mp(m + \textit{M} H))C_{1,3}-\sqrt{4n\gamma}C_{4,2}-p_3C_{3,1}=0$$
\begin{equation}
(\varepsilon \mp(m - \textit{M} H))C_{2,4}-\sqrt{4n\gamma}C_{3,1}+p_3C_{4,2}=0 .
\end{equation} \\
The condition of equality of determinant of system to zero leads as to anexpression for energy spectrum  $\varepsilon^{(n)}$ (see also [5]) as following
\begin{equation}
\varepsilon^{(n)}=\sqrt{p_3^2+(\sqrt{m^2+4n\gamma}+s\textit{M} H)^2} \  ,  \ \ \ \ \ \  s=\pm 1.
\end{equation} 

It is possible to show that  $\hat{H}_D$ commutates with the operator  $\hat{\mu}_3$ [7]:
\begin{equation}
\hat{\mu}_3=m\Sigma_3+\rho_2\left[\vec{\Sigma}\times\vec{P}\right]_3 \  , \ \ \ \  \rho_2= \left(\begin{array}{cc}
0 & -iI \\
iI & 0
\end{array}\right) ,
\end{equation} \\
which describes the spin states of the fermion, 
the projection of the spin of proton on the direction of the magnetic field. Therefore
 $$\hat{\mu}_3\psi=K_0\psi, $$ 
 where $K_0$ the eigenvalue of $\hat{\mu}_3$.
 
 It follows from the last equation that the coefficients of $C_i$ satisfy the system of the linear equations:
$$(\varepsilon^{(n)}-K_0- \textit{M} H)C_1=+p_3C_3$$
$$(\varepsilon^{(n)}+K_0+ \textit{M} H)C_2=-p_3C_4$$
\begin{equation}
(\varepsilon^{(n)}+K_0+ \textit{M} H)C_3=+p_3C_1
\end{equation}
$$(\varepsilon^{(n)}-K_0- \textit{M} H)C_4=-p_3C_2 .$$ \\
during obtaining of this system we used the easily checked relationship:
$$\hat{\mu}_3\psi=(\varepsilon^{(n)}\rho_3\Sigma_3-i\rho_2p_3)\psi .$$
  From system of equ.(9) we obtain expression for the eigenvalue of the operator $\hat{\mu}_3$ :
\begin{equation}
K_0=-\textit{M} H+s\sqrt{\varepsilon^{(n)}{^2}-p_3^2} \  , \ \ \ \ \ \  s=\pm1 .
\end{equation} 

Coefficients  $C_i$ can be found by deciding together equ.(6) and equ.(9). In this case it occurs that the solution exists only if the signs of $s$ in relations (7) and (10) are opposite. Thus the following agreement rule of signs is obtained:
$$\varepsilon^{(n)}=\sqrt{p_3^2+(\sqrt{m^2+4n\gamma}+s\textit{M} H)^2}$$
\begin{equation}
K_0=-\textit{M} H-s\sqrt{\varepsilon^{(n)}{^2}-p_3^2} \ , \ \ \ \ \ \  s=\pm1.
\end{equation}

Takeing into account normalization condition:
\begin{equation}
\sum_{i=1}^{4}\left|C_i\right|^2=1 .
\end{equation}
for $C_i$ we obtain :
\begin{equation}
\left[\begin{array}{c}
C_1 \\
C_2 \\
C_3 \\
C_4
\end{array}\right]=(1+A^2)^{-\frac{1}{2}}(1+B^2)^{-\frac{1}{2}}\left[\begin{array}{c}
1 \\
-AB \\
A \\
B
\end{array}\right]
\end{equation}  
where ,
$$A=\frac{\varepsilon^{(n)}-K_0-\textit{M} H}{p_3}  ,  \ \ \ B=-\frac{K_0+m+2\textit{M} H}{\sqrt{4n\gamma}} .$$ \\
 Note that these coefficients do not go to infinity with  $p_3= 0  ,  n=0$.
 
In the case of the neutral particle (neutron) the wave function and energy spectrum take, correspondingly, the form:
\begin{equation}
\psi_n(r)=\frac{1}{L^{{3}/{2}}}e^{i(p_1 x+p_2 y+p_3 z)}
\left[\begin{array}{c}
C_1 \\
C_2 \\
C_3 \\
C_4
\end{array}\right]_n
\end{equation}

\begin{equation}
\varepsilon_{n}^{(p_\bot)}=\sqrt{p_3^2+(\sqrt{m_n^2+p_\bot^2}+s\textit{M}_n H)^2} \ , \ \  p_1^2+p_2^2=p_\bot^2 \; , \ \ \  s=\pm1\;,
\end{equation} \\
where $m_n$ - is mass , $\textit{M}_n$ - is AMM of neutron.

The eigenvalues of the operator  $\hat{\mu}_3$ (with $e=0$) are determined by formula (10) with the rule of the agreement of signs equ. (11). The coefficients of $C_i$ are determined by formula (13), in which $B$ takes the form:
\begin{equation}
B=-\frac{K_{0n}+m_n+2\textit{M}_n H}{p_\bot}.
\end{equation} \\

\begin{center}
\title\textbf{3. Total probability near the threshold}
\maketitle
\end{center}

The inverse decay process of proton is described by Lagrangian:
\begin{equation}
L=\frac{G_F}{\sqrt2}\left[\bar{\Psi}_n\gamma_\mu(1+\alpha\gamma_5)\Psi_p\right]\cdot\left[\bar{\Psi}_\nu\gamma^\mu(1+\gamma_5)\Psi_{e^+}\right],
\end{equation} \\
where $G_F$ - is Fermi constant, $\alpha $ - is the relation of the constants of the axial-vector and vector interactions of $G_A$ and $G_V$ ($\alpha\approx 1.25$), $\gamma_\mu$, $\gamma_5$ - are Dirac's matrices.\\ 
 We will consider process near the threshold, i.e. when:
$$\Delta=m- \textit{M} H-(m_n-\textit{M}_n H)-m_e<<eH\leq m_e .$$
Here $m_e$ - is mass of positron.

The wave function of positron can be put from the wave function of proton, if we replace $m$ by $m_e$ and to place $\textit{m}=0$, since in contrast to (kinematic) anomalous magnetic moments of the nucleons, positron  has dynamic nature AMM, and in the energy range of values of the external magnetic field, it is the disappearing function of the magnetic field (for example, see [9]).

The wave function of neutrino can be obtained in the form:
$$\Psi_{\nu}(\vec r , t)=\frac{1}{2L^{{3}/{2}}}e^{-i\varepsilon t+i\vec{p}\cdot\vec{r}}\left[\begin{array}{c}
f_1 \\
f_2 \\
-f_1 \\
-f_2
\end{array}\right],$$ \\где 
\begin{equation}
f_2=(1+\frac{p_3}{\left|\vec{p}\right|})^{1/2}  , \ \ \  f_1=-e^{-i arctg{\frac{p_y}{p_x}}}(1-\frac{p_3}{\left|\vec{p}\right|})^{1/2} .
\end{equation}

We conduct calculations in the frame of reference, where the proton  is unmoved \\($p_3=0 ,  n=0$). Since we consider process near the threshold, in the first approximation, it is possible to disregard momentum of neutron in comparison with $m_n$ in the coefficients  $C_i$ of wave function. Then, after integration for the space coordinates of all four particles, for the square of the module of matrix element we obtain the following expression:
\begin{equation}
\left|M\right|^2=\frac{G^2_F}{4L^{10}}{(1-\alpha)}^2 {(1+\frac{p_{3\nu}}{\left|\bar{p}_{\nu}\right|})}\cdot\frac{{(1+C)}^2}{1+C^2}\cdot \exp{\left[-\frac{p^2_{2e^+}+(p_{1n}+p_{1\nu})^2}{4\gamma}\right]},
\end{equation} 
where  \\ $$C=\frac{\sqrt{m^2+p^2_{3e^+}}-m}{p_{3e^+}}.$$ 

The total probability of the inverse  $\beta$ decay of proton is determined by the following expression:
$$W=\frac{L^{10}}{{(2\pi)}^5}\int{\cdots}\int{\left|M\right|^2}\delta{(\varepsilon_p-\varepsilon_n-\varepsilon_{e^+}-\varepsilon_\nu)}\cdot\delta{(p_{3n}+p_{3e^+}+p_{3\nu})}\cdot$$
\begin{equation}
\cdot\delta{(p_{2n}+p_{2e^+}+p_{2\nu})}\cdot d^3p_n d^3p_\nu dp_{3e^+} dp_{2e^+}.
\end{equation} \\
We considered that near the threshold the contribution to the sum on $n$ will give only term $n=0$. first with the aid of $\delta$ - function on $p_2$ and $p_3$ let us integrate over the momentums of positron, and then with the aid of $\delta$ - function on energy will find the range of changes in the momentums of neutron and neutrino:
$$\left|p_{1n}\right|\leq\sqrt{2m_n\Delta}  ;   \left|p_{1\nu}\right|\leq\Delta$$
\begin{equation}
\left|p_{2n}\right|\leq\sqrt{2m_n\Delta}  ;   \left|p_{2\nu}\right|\leq\Delta
\end{equation}
$$\left|p_{3n}\right|\leq\sqrt{2m_n\Delta}  ;   \left|p_{3\nu}\right|\leq\Delta.$$ 

Hence it follows that taking into account the smallness of $\Delta$ everywhere in the integral it is possible to disregard the momentum of neutrino in comparison with the momentum of neutron. Further, passing to the cylindrical coordinates
$$d^3p_n d^3p_\nu=(2\pi)^2p_{\bot n}dp_{\bot n}p_{\bot\nu}dp_{\bot\nu} dp_{3n}dp_{3\nu}.$$
the dominant term of the probability of the process in question we obtain in the form:
\begin{equation}
W\cong\frac{G^2_F}{32\pi^3}\sqrt{2}\;(1-\alpha)^2 \cdot m_e^{5} \cdot \widetilde{\Delta}^{5/2},
\end{equation}
where - $\widetilde{\Delta}=\Delta/m_e$.\\
  Thus, total decay probability proves to be proportional: $\widetilde{\Delta}^{5/2}$.\\
         
Note that the analogs URCA - processes in the quarks substance are the electro-weak decay  $\underline{u}$ and $\underline{d}$  quarks. These decays can be also described by Lagrangian (17), in which the wave functions of proton and neutron must be replaced, correspondingly, with the wave functions of $\underline{u}$ and
$\underline{d}$  quarks and, furthermore, to place the parameter  $\alpha=1$. If quarks possess kinematic AMM, then for the decay probability of $\underline{u}$  quark in the strong magnetic field it is possible to use formula (22), in which should be placed $\alpha=1$.

Then the dominant term of the total decay probability  $\underline{u}$ quark with AMM in the strong magnetic field is reduced to zero. Consequently, in this approximation  $(\propto\widetilde{\Delta}^{5/2})$ the electro-weak decay of $\underline{u}$  quark with AMM in the lowest energy state proves to be completely (kinematically) forbidden.\\
\\

At the end the authors express their appreciations to Prof. Khalilov V. R. for the formulation of the problem, aid in conducting of calculations, and also a whole series of useful observations.\\
\\
\\

\end{document}